\documentclass[]{aa}
\usepackage{graphicx,natbib}
\newcommand{\lya}{Ly$\alpha$}
\newcommand{\ha}{H$\alpha$}
\newcommand{\nii}{[\ion{N}{ii}]}
\newcommand{\ecs}{erg\,cm$^{-2}$\,s$^{-1}$}
\newcommand{\kms}{km\,s$^{-1}$}

\newcommand{\rg}{PKS~1138$-$262}
\newcommand{\rgs}{1138$-$262}
\newcommand{\p}{$\pm$}
\begin{document}
  \title{A search for clusters at high redshift}
  \subtitle{IV. Spectroscopy of \ha\ emitters in a proto-cluster at
  $z=2.16$}

  \author{J.~D. Kurk \inst{1} \and L. Pentericci \inst{2} \and
          R.~A. Overzier \inst{1} \and H.~J.~A. R\"ottgering \inst{1} \and 
          G.~K. Miley \inst{1}}

  \institute{Sterrewacht Leiden, P.O. Box 9513, 2300 RA, Leiden, 
             The Netherlands               
        \and
             Max-Planck-Institut f\"ur Astronomie, K\"onigstuhl 17,
             D-69117, Heidelberg, Germany}

%   \offprints{J.D. Kurk, \email{kurk@strw.leidenuniv.nl}}
%  \offprints{J.D. Kurk (kurk@strw.leidenuniv.nl)}

  \date{Received date / Accepted date}

\abstract{ 
Radio galaxy \object{PKS~1138$-$262} is a massive galaxy at $z =
2.16$, located in a dense environment. We have found an overdensity of
\lya\ emitting galaxies in this field, consistent with a proto-cluster
structure associated with the radio galaxy.  Recently, we have
discovered forty candidate \ha\ emitters by their excess near infrared
narrow band flux.  Here, we present infrared spectroscopy of
nine of the brightest candidate \ha\ emitters.  All these candidates
show an emission line at the expected wavelength.  The identification
of three of these lines with \ha\ is confirmed by accompanying
[\ion{N}{ii}] emission. The spectra of the other candidates are
consistent with \ha\ emission at $z \sim 2.15$, one being a QSO as
indicated by the broadness of its emission line.  The velocity
dispersion of the emitters (360 km s$^{-1}$) is significantly smaller
than that of the narrow band filter used for their selection (1600 km
s$^{-1}$).  We therefore conclude that the emitters are associated
with the radio galaxy.  The star formation rates (SFRs) deduced from
the \ha\ flux are in the range 6 -- 44 M$_\odot$ yr$^{-1}$ and the SFR
density observed is 5--10$\times$ higher than in the HDF-N at $z =
2.23$.  The properties of the narrow emission lines indicate that the
emitters are powered by star formation and contain very young ($< 100$
Myr) stellar populations with moderately high metallicities.

 }

\maketitle
%
%________________________________________________________________

\section{Introduction}
High redshift clusters are prime subjects for the study of galaxy
formation and cosmology. The powerful radio galaxy \rg\ at $z = 2.156$
appears to be the brightest galaxy in a high redshift cluster. We have
discovered an overdensity of \lya\ emitters within 1.5 Mpc of \rgs\
\citep[][Paper I]{kur00}. The redshifts of 14 emitters were
spectroscopically confirmed to be in the range $2.14 < z < 2.18$
\citep[][Paper II]{pen00a}. In addition, we have carried out near
infrared imaging \citep[][Paper III]{kur03a}. The number of $K$ band
galaxies and extremely red objects in this field is higher than in
blank fields.  We found 40 objects with excess narrow band flux,
consistent with \ha\ emission at $z \sim 2.16$. The surface density of
\ha\ emitters increases towards the radio galaxy and their average $K$
magnitude is lower and therefore their inferred stellar mass higher
than for the \lya\ emitters.  Here, we present infrared spectroscopy
of nine candidate \ha\ emitters to confirm their redshift and
determine the velocity dispersion of the sample.  We assume a flat
Universe with $h_0=0.65$ and $\Omega_{\rm m}=0.3$.

%__________________________________________________________________

\section{Selection, observations and data reduction}\label{obs}
With the aim of detecting \ha\ emitting galaxies in the proto-cluster
associated with \rg, imaging in $K_s$ and in a narrow band filter
($\lambda_{\mathrm c} = 2.07\,\mu$, FWHM = 0.026$\,\mu$) was carried
out, employing two pointings covering a total field of 12.5
arcmin$^2$. There are 40 objects with rest frame equivalent width
(EW$_0$) $> 25$ \AA\ (see Paper III). From the list of 29 candidate
\ha\ emitters within 1\farcm3 from the radio galaxy, we
selected those with \ha\ flux $> 3.5 \times 10^{-17}$
\ecs. Furthermore, we selected those that were conveniently located
for placement in the slit for spectroscopic follow-up, which was
carried out with ISAAC at VLT Antu (UT1)\footnote{Based on
observations carried out at the European Southern Observatory,
Paranal, Chile, project P68.A-0184(A).}.  The short wavelength camera
of ISAAC is equipped with a Rockwell Hawaii 1024$^2$ pixel Hg:Cd:Te
array which has a projected pixel scale of 0.147\arcsec. We used the
medium resolution grating in second order resulting in a dispersion of
1.23\,\AA. The observations were carried out in the nights of March 23
and 25, 2002 under variable seeing, which was just below 1\arcsec\ for
most of the time. The 1\arcsec$\times$120\arcsec\ slit employed
resulted in a resolution of $\sim 2600$. During acquisition, the slit
was first positioned on a bright point source (a star or the radio
galaxy) and subsequently positioned at the midpoint between two
candidates, which was always within 32\arcsec. We have employed four
slit positions, each targeted at two or three candidate emitters for
3.5 hours (3.75 in one case). In total, nine candidates were observed,
one of which was included in two slits. Per slit, we obtained 14 (15)
frames of 15 minutes with offsets of 15, 18 or 20\arcsec\ in ABBA
sequence with additional random jitter offsets up to 5, 2, or
1\arcsec, respectively, to avoid recurrent registration of spectra on
bad pixels. Standard stars were observed with the same slit at a range
of airmasses during the nights to correct for telluric absorption and
to calibrate the data in flux. All observations were carried out at
airmass below 1.8.

%Data reduction was done in IRAF, Eclipse and IDL\footnote{IRAF stands
%for Image Reduction and Analysis Facility and is distributed by the
%National Optical Astronomy Observatories. Eclipse is a collection of
%programs used in the ISAAC pipeline reduction software (Devillard, ESO
%Messenger 87, 1997). IDL is the Interactive Data Language developed by
%Research Systems, Inc.}. First, from all frames ghosts of bright
%spectra were subtracted. Subsequently, they were flat fielded with the
%sky flat for the respective night. The ISAAC spectroscopy pipeline was
%employed to subtract the frames pairwise, which yields good sky
%(lines) and bias subtraction and to correct for field
%distortion. Offsets between images were determined from the instrument
%setup parameters. The frames were combined by computing the median of
%each pixel stack to reject cosmic rays and bad pixels, resulting in a
%final frame with positive spectra accompanied by negative spectra on
%both sides.

\begin{figure*}
  \centerline{
  \resizebox{0.7\hsize}{!}{\includegraphics{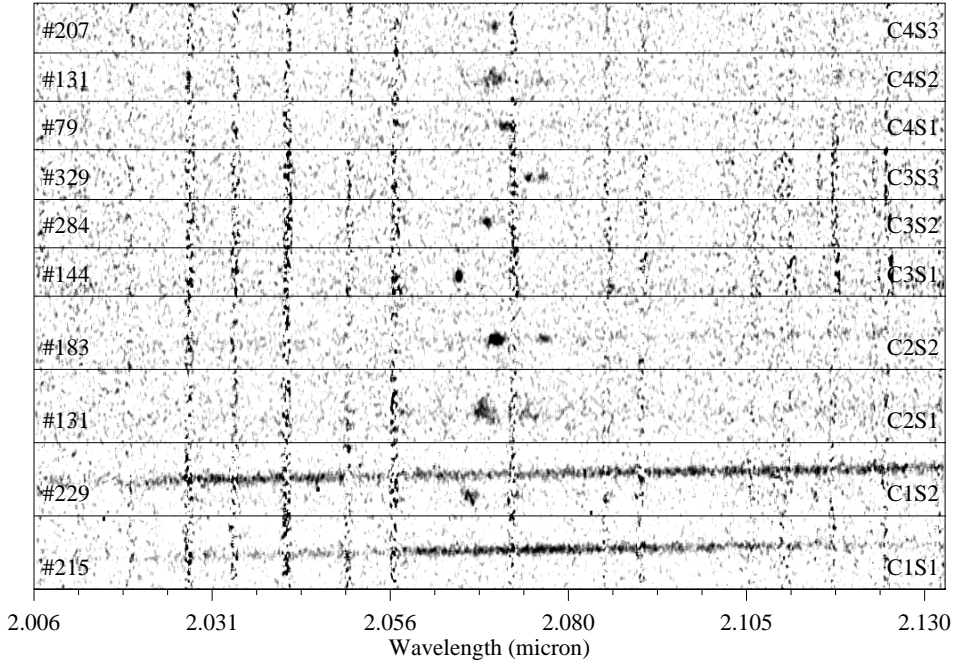}}
  }
  \caption{The ten two dimensional spectra observed through four
  slits. Skyline residuals are visible as vertical lines with a higher
  noise level. Slit and spectrum number are indicated, as are the
  object number from the \ha\ candidate list.  The images have been
  smoothed with a mask of 3$\times$3 pixels.  Near the emission line
  of candidate 229, continuum emission of a serendipitous galaxy is
  visible. The spectrum of candidate 215 shows a very broad line which
  almost covers the full spectral range.}
  \label{twod}
\end{figure*}

Standard data reduction was carried out using pairwise frame
subtraction, resulting in a final frame with positive spectra
accompanied by negative spectra on both sides. Care was taken during
setup to ensure that these negative spectra did not overlap with
positive ones. One dimensional spectra of the candidate emitters were
extracted from the positive two-dimensional spectra using the spatial
profile of a standard star spectrum observed during that night and
averaged per 3 pixels yielding bins of 3.7\,\AA. The spectral resolution
is 7\,\AA, as measured from the FWHM of the skylines. The wavelength
calibration is based on the OH skylines observed. The telluric
standards observed at a range of airmasses show only small variations
(less than a percent on average). An average absorption spectrum per
night was used to correct for telluric absorption. There are no
spectrophotometric flux standards in the infrared. We have therefore
used one of the telluric standards (Hip043868) with spectral type
B1. Such stars have a featureless spectrum in this wavelength region
given by a black body curve at T = 25,500 K. The curve was normalized
to the $K$ magnitude of the star and subsequently used to divide the
extracted and absorption corrected spectrum to obtain the flux
calibration. The two dimensional spectra of all objects are shown in
Fig.\ \ref{twod}.

\section{Results}\label{results}

\begin{figure}
  \resizebox{\hsize}{!}{\includegraphics{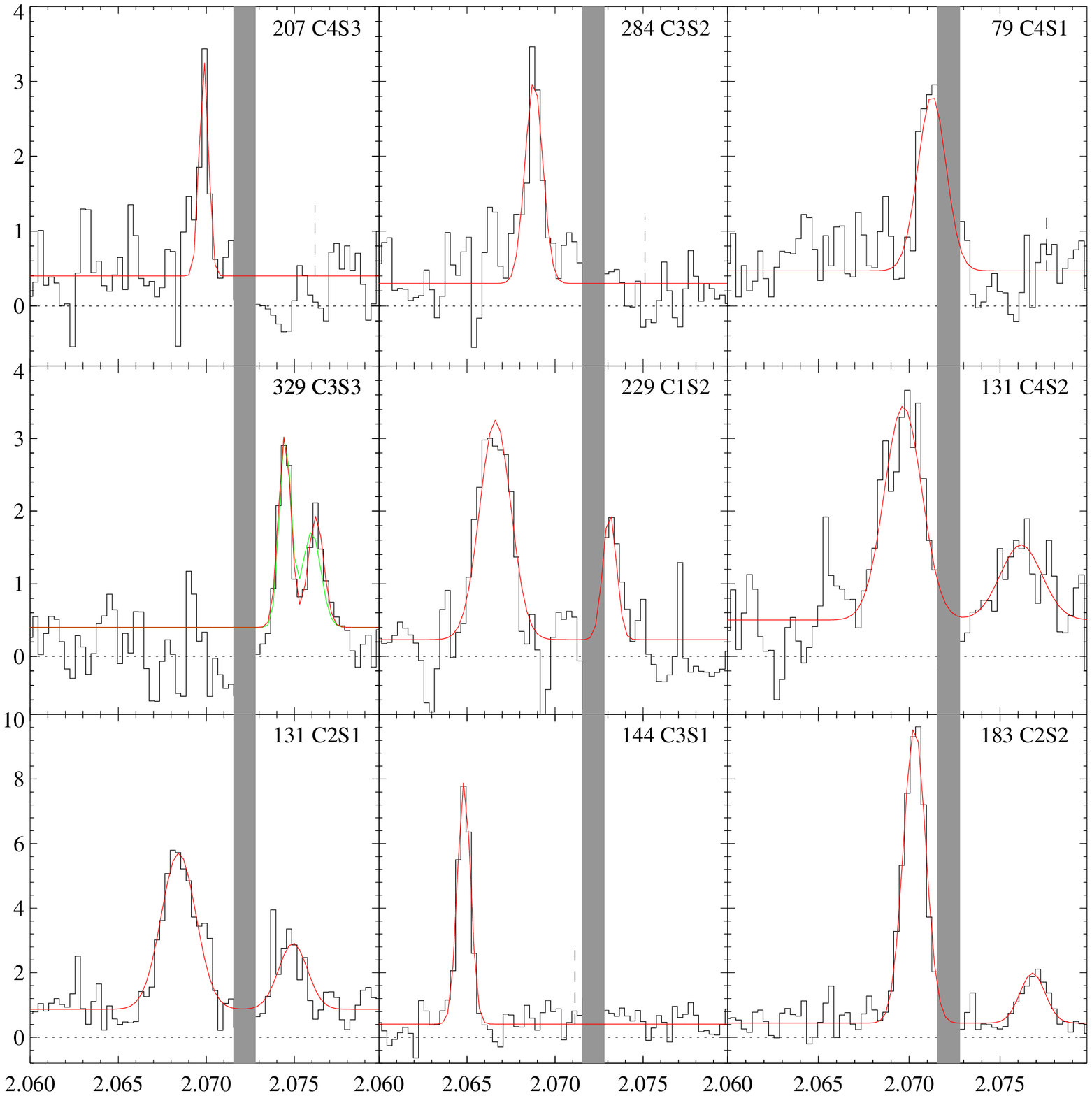}} \caption{One
  dimensional spectra of the eight narrow line emitters (histograms)
  with fits overlayed (solid lines).  Object 329 has also an
  alternative fit for [\ion{O}{II}].  Units are in 10$^{-18}$
  \ecs\,\AA$^{-1}$ and $\mu$m.  The grey band denotes the position of
  a sky line.  The dashed line indicates the position of the
  [\ion{N}{II}] line for identification of the detected line with
  \ha. Note the deviant flux density scale in the last row.}
  \label{fits3x3}
\end{figure}

All candidates observed show an emission line, which means that our
selection was 100\% efficient. We have fit Gaussian curves to the
emission lines applying a least squares method (see Fig.\
\ref{fits3x3}) in order to determine their central wavelength,
deconvolved FWHM (if resolved), flux and EW$_0$ (Table \ref{lines}).
Also presented in this table is the SFR derived from the \ha\ and UV
luminosities (see discussion in Paper III), which is in the range 6 --
44 M$_\odot$ yr$^{-1}$.  We do not detect continuum emission in most
of the spectra, except in the co-added spectrum of candidate 131, 215
and the serendipitous object in slit one (0.6$\pm$0.6, 0.7$\pm$0.6 and
1.7$\pm$0.7 $\times 10^{-18}$ \ecs\,\AA$^{-1}$, respectively), which
compare favourably with the line subtracted broad band fluxes measured
by imaging in paper III (0.6, 1.0 and 2.1 $\times 10^{-18}$
\ecs\,\AA$^{-1}$, respectively).  The EW$_0$ is therefore based on the
line flux measured in the spectra and the broad band magnitude
measured on the images.  

The spectra of objects 131, 183 and 229 were fit by two Gaussians for
which the relative centers were fixed as for \ha\ and
\nii$\lambda6583$\AA.  Candidate 329 shows two emission peaks only
17.8$\pm$2.7\,\AA\ apart. Both lines could be \ha\ emission from one
galaxy with two components, seperated by $\sim 250$ \kms\ in velocity.
We have considered the possibility that the lines are due to
[\ion{O}{ii}] emission at $z = 4.5674$ which would have a separation
of 15.0\,\AA.  The emission line ratio of the supposed
[\ion{O}{ii}]\,$\lambda3729/3726$ would be 0.7 which implies an
electron density $\sim 10^3$ cm$^{-3}$, normally only observed in the
central parts of nebulae \citep{ost89}.  In addition, a faint ($B =
27.0$) counterpart in the $B$ band, sampling a wavelength range below
912\,\AA\ for $z = 4.6$, makes the identification with [\ion{O}{ii}]
improbable.  Candidate 131 was included in two slits, with a position
angle difference of 40$^\circ$.  The two fits to the spectra of
emitter 131 indicate a velocity difference of 180 \kms\ which can be
explained by the fact that different regions of the galaxy have been
sampled.

For how many of the emission lines can we be sure that the
identification with \ha\ at $z \sim 2.15$ is correct?  For the three
objects with confirming \nii\ lines, we can be certain. For the QSO
object 215, \nii\ is blended with the very broad \ha\ line and
impossible to discern. Given a \nii/\ha\ line ratio of 1/3, we do not
expect to detect the \nii\ line for objects 79, 207, 284, 329 above
the noise and the identification with \ha\ is therefore consistent but
not 100\% certain. An identification with \ha\ for object 144 can only
be true if the \nii/\ha\ ratio is $< 1/6$ (object 183 has an observed
ratio of 1/5.5). An alternative identification with
[\ion{O}{iii}]$\lambda$5007\AA\ is improbable, as we do not detect its
counterpart [\ion{O}{iii}]$\lambda$4958\AA\ at 2.045$\mu$. We consider
this therefore a probable \ha\ identification.

At least three of the emission lines are spatially extended. In two
cases, we detect unordered velocity structure, but the morphology of
the two dimensional spectrum of candidate 229 resembles a rotation
curve.  A fit to this structure results in a rotation velocity of 50
\kms\ at 6 kpc radius, implying a dynamical mass of $3.5 \times
10^{9}$ M$_\odot$.

\begin{table*}
\begin{center}
\caption[]{Properties of the observed emission lines\vspace{-0.2cm}}
\hspace{-1mm}\begin{tabular}{r c c l c r r @{\p}r r @{\p}r r @{\p}r r @{\p}r r @{\p}r r}
\hline \hline \label{lines}
ID & S & R.A. \& Dec. (J2000) & Line & Fit &
\multicolumn{1}{c}{$z$} & \multicolumn{2}{c}{FWHM} &
\multicolumn{2}{c}{Flux} & \multicolumn{2}{c}{EW$_0$} & 
\multicolumn{4}{c}{$_\mathrm{H\alpha}$\,SFR\,$_\mathrm{UV}$} & O/H \\
(1) & (2) & (3) & \multicolumn{1}{c}{(4)} & (5,6) & 
\multicolumn{1}{c}{(7)} & \multicolumn{2}{c}{(8)} & \multicolumn{2}{c}{(9)} & 
\multicolumn{2}{c}{(10)} & \multicolumn{2}{c}{(11)} & \multicolumn{2}{c}{(12)} & (13) \\ \hline
 79 & 4 & 11:40:52.62 $-$26:30:01.0 & \ha & 1,2 & 2.1558 & 160 &  80 &  3.6 
& 1.9 &  90 &  50 & 11 & 6 & 33 & 1       \\
131 & 2 & 11:40:51.28 $-$26:29:38.7 & \ha & 3,1 & 2.1518 & 320 &  50 & 12.3 
& 2.2 &  65 &  15 & 39 & 7 & 19 & 1 & 8.8 \\
    &   &                         & \nii&     &        & 270 & 100 &  4.4 
& 2.0 &  25 &  10 & \multicolumn{2}{c}{} & \multicolumn{2}{c}{}   &\p0.2\\
131 & 4 & 11:40:51.28 $-$26:29:38.7 & \ha & 3,1 & 2.1537 & 360 &  80 &  8.2 
& 2.4 &  45 &  15 & 26 & 8 & 19 & 1 & 8.8 \\
    &   &                         & \nii&     &        & 420 & 250 &  3.4 
& 2.5 &  20 &  15 & \multicolumn{2}{c}{} & \multicolumn{2}{c}{}   &\p0.2\\
144 & 3 & 11:40:43.45 $-$26:29:37.5 & \ha & 1,3 & 2.1463 &  40 &  20 &  6.7 
& 1.3 & 330 &  70 & 21 & 4 & 15 & 1       \\
183 & 2 & 11:40:46.15 $-$26:29:24.9 & \ha & 3,1 & 2.1546 & 170 &  20 & 13.8 
& 1.7 & 230 &  35 & 44 & 5 & 40 & 1 & 8.6 \\
    &   &                         & \nii&     &        & 200 & 120 &  2.5 
& 1.7 &  40 &  30 & \multicolumn{2}{c}{} & \multicolumn{2}{c}{}   &\p0.4\\
207 & 4 & 11:40:50.20 $-$26:29:21.0 & \ha & 1,2 & 2.1540 & 
                               \multicolumn{2}{l}{\hspace{-0.9mm}$<$100}
& 1.9 & 0.9 & 50 &  25 &  6 & 3  & 12 & 1 \\
215 & 1 & 11:40:46.01 $-$26:29:16.9 & \ha & 3,1 & 2.1568 & 5300 & 800 & 46.2
& 8.8 & 150 &  30 & \multicolumn{2}{c}{$^\dagger$}\\
229 & 1 & 11:40:46.10 $-$26:29:11.5 & \ha & 3,1 & 2.1489 &  290 &  60 &  7.1 
& 1.9 &  30 &  10 & 23 & 6 & 26 & 1 & 8.8 \\
    &   &                         & \nii&     &        &  130 &  90 &  2.4 
& 1.7 &  11 &   8 & \multicolumn{2}{c}{} & \multicolumn{2}{c}{}   &\p0.3\\
284 & 3 & 11:40:45.58 $-$26:29:02.4 & \ha & 1,2 & 2.1523 &   90 &  50 &  3.0
& 1.4 & 300 & 140 & 10 & 4 & 6 & 1        \\
329 & 3 & 11:40:46.88 $-$26:28:41.4 & \ha & 2,3 & 2.1609 &   60 &  40 &  2.7
& 1.3 &1350 & 700 &  9 & 4 & 7 & 1 \\
    &   &                         & \ha & 2,3 & 2.1636 &  110 &  30 &  2.0
& 1.1 &1000 & 550 &  6 & 4 \\
\hline \hline
\end{tabular}
\end{center}\vspace{-0.1cm}
%\begin{center}
\footnotesize \noindent Notes: (1) Candidate number (2) Slit number
(3) Coordinates (4) Line identification (5) Type of fit: 1) one
Gaussian curve, 2) two Gaussian curves, 3) two Gaussians for \ha\ and
\nii\ (6) Identification is 1) certain, detection of \nii, 2)
consistent, expected non detection of \nii, 3) possible (7) Redshift
with random error 0.0002, except for candidate 215, for which it is
0.002 (8) FWHM in km s$^{-1}$ (9) Flux in 10$^{-17}$ \ecs\ (10) EW$_0$
in \AA\ (11) SFR$_\mathrm{H\alpha}$ in M$_\odot$ yr$^{-1}$ (12)
SFR$_\mathrm{UV}$ in M$_\odot$ yr$^{-1}$ (13) Metallicity in 12 +
log[O/H] ($^\dagger$) \ha\ unrelated to SFR.
%\end{center}
\vspace{-0.2cm}
\end{table*}                   

\section{Discussion}\label{discussion}

\begin{figure}
  \resizebox{\hsize}{!}{\includegraphics{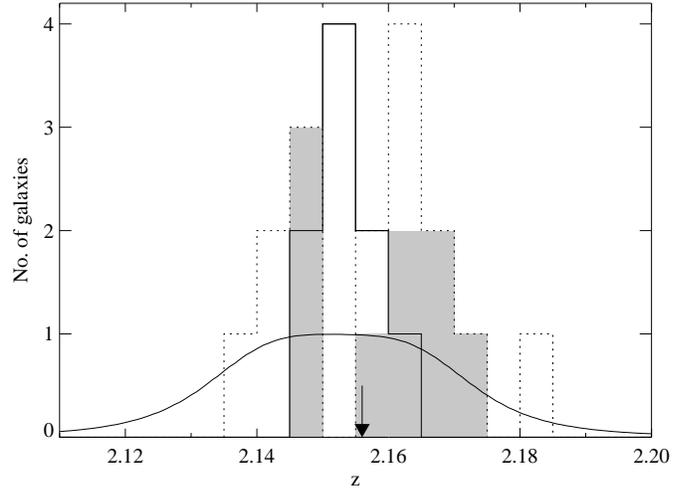}}
  \caption{Redshift histogram for the \ha\ emitters (solid) and the
  previously known \lya\ emitters (dotted, and shaded for those within
  ISAAC field). An arrow denotes the redshift of \rgs. Also shown is
  the sensitivity curve of the narrow band filter used for the
  selection of the \ha\ candidates. The surface beneath the curve is
  normalized to nine.}
  \label{zhist}
\end{figure}

In Fig.\ \ref{zhist} we show the redshift distribution of the 9
emitters, assuming that all emission lines can be identified as \ha.
For this plot, we used the average redshift of the two lines of object
329. Also plotted is the sensitivity curve of the narrow band filter
used to select the candidates. A random distribution of emitter
redshifts would follow this curve. We have run a Monte Carlo
simulation with 10,000 realizations of a randomly sampled distribution
of emitters given this filter as selection criterium. Although the
mean of the measured distribution is only 0.18$\sigma$ away from the
mean of a random sample, its dispersion is much smaller, deviating by
1.75$\sigma$ from a random sample. The probability that the redshifts
we measure are drawn from a random distribution is therefore 8\%. Note
that the redshift of the radio galaxy is 2.156 \citep{rot97}, very
close to the mean of the selection filter (2.152), and to the mean of
the measured redshift distribution (2.154). The distribution is
consistent with a group of \ha\ emitters associated with the radio
galaxy. The velocity dispersion of this group is 360 \kms\
\citep[using the gapper sigma method,][]{bee90}, while the virial
radius of the nine emitters is 0.45 Mpc, implying a virial mass of
$8\times10^{13}$ M$_\odot$ (assuming that all lines can be identified
with \ha\ and taking the mean of the two redshifts for object 131).
This mass is merely illustrative as at this redshift it is improbable
that the structure is virialized.  The velocity dispersion of the \ha\
emitters is smaller than the velocity dispersion of the confirmed
\lya\ emitters, both for the complete sample (1050 \kms) and the nine
within the solid angle of the two ISAAC fields (760 \kms, shaded part
of histogram in Fig.\ \ref{zhist}).  There is no evidence for a
bimodal redshift distribution as observed for the \lya\ emitters.

We can construct a complete sample out of the spectroscopic sample by
excluding the two objects with the lowest \ha\ flux and including the
radio galaxy.  This collection represents all candidate \ha\ emitters
with F$_{\rm H\alpha} > 4.0 \times 10^{-17}$ \ecs\ within 1\farcm3
from the radio galaxy. The FWHM of the narrow band filter ($2.134 < z
< 2.174$) and the solid angle given above define a comoving volume of
815 Mpc$^3$, resulting in a volume density of 0.010 Mpc$^{-3}$, which
is a factor four higher than the density of confirmed \lya\ emitters
in this field.  All star forming objects detected have line fluxes
lower than the high redshift \ha\ surveys discussed in Paper III, but
we can compare the SFR density to the density at $z = 2.23$ derived
from \ha\ emission in the HDF-N as measured by \citet{iwa00}.
Following their cosmology ($h_0=0.5$, $q_0=0.5$) and procedure to
correct for the part of the \ha\ luminosity function below the
detection limit, we obtain a SFR density of 0.48 M$_\odot$ yr$^{-1}$
Mpc$^{-3}$.  This is 10 (5) times higher than the (reddening
corrected) value obtained for the HDF-N. Using the redshift range
defined by the \ha\ emitters ($2.146 < z < 2.164$) results in values
that are larger by a factor of two.  Likewise, smaller SFR densities
would result if some of the \ha\ lines have been misidentified.

The properties of the detected emission lines provide information
about the physical conditions in the galaxies.  The FWHM of the narrow
nebular emission lines detected are in the range 40 -- 360 \kms\ with
an average of 190 \kms. These values are comparable to those found for
LBGs at $z \sim 3$ by \citet{pet01}.  The [\ion{N}{ii}]/\ha\ ratio can
be used to distinguish narrow-line active galaxies from \ion{H}{ii}
region-like galaxies \citep{vei87}.  The three emitters with detected
[\ion{N}{ii}] have $-0.74 <$ log([\ion{N}{ii}]/\ha) $< -0.42$, which
puts them among the star forming galaxies.  In the absence of shock
excitation, the [\ion{N}{ii}]/\ha\ ratio can also be used as
metallicity indicator.  Using the empirical relation calibrated by
\citet{den02}, the average ratio of the three emitters implies 12 +
log(O/H) $\approx$ 8.7. This value is comparable to the broad range of
values obtained for present-day spiral galaxies \citep{zee98}.  The
EW$_0$ of some detected narrow lines are surprisingly high, up to
1350\,\AA. This can be explained by very young stellar populations
where the continuum radiation around 6000\,\AA\ is still very weak.
EW$_0$ values between 200 and 330\,\AA\ imply an age $< 100$ Myr
\citep{lei99}. The moderately high metallicities found for the objects
with detected [\ion{N}{ii}] emission, however, require that the
galaxies are near the end of the star formation event.  This
requirement seems to indicate that these emitters have undergone a
very similar evolution.

\section{Conclusion}\label{conclusion}
Infrared spectroscopy has established the presence of nine line
emitters within 0.6 Mpc of the HzRG \rg. Three emitters show an
additional line which confirms the identification with \ha\ at $z =
2.15$, while four more have spectra consistent with \ha\ at this
redshift, one being a QSO as indicated by the broadness of its
emission line. One emitter shows only a single strong line, which is
possibly \ha\ and one emitter exhibits two lines which probably
originate from two emission line regions within one galaxy at $z =
2.16$. Additional evidence for identification of all observed lines
with \ha\ is the small velocity dispersion (360 km s$^{-1}$) as
compared with the width of the selection filter. This dispersion is
also smaller than the dispersion of the \lya\ emitters. The star
formation rate density of the observed emitters is a factor ten higher
than found at $z = 2.23$ in the HDF-N.  These results support the
formerly advocated ideas that \rg\ is located in a proto-cluster at $z
= 2.16$.  The properties of the narrow emission lines indicate that
the emitters are powered by star formation and contain very young ($<
100$ Myr) stellar populations with moderately high metallicities. It
seems that we observe these galaxies near the end of their first and
major burst of star formation.

\begin{acknowledgements}
We are grateful to the ESO VLT staff for excellent support during the
observing run.  We acknowledge fruitful discussions with B.\ Venemans
and S.\ di Serego Alighieri.  Comments of the anonymous referee have
also helped to improve the manuscript.  This research has made use of
the NASA/IPAC Extragalactic Database (NED) which is operated by the
Jet Propulsion Laboratory, California Institute of Technology, under
contract with the National Aeronautics and Space Administration. We
have also made use of NASA's Astrophysics Data System Bibliographic
Services.
\end{acknowledgements}

\bibliographystyle{aa}
\bibliography{aa1819.bib}

\end{document}